\documentstyle[aps,prd,floats,epsfig]{revtex}
\begin{document} 
\twocolumn[\hsize\textwidth\columnwidth\hsize\csname
@twocolumnfalse\endcsname
\draft
\title{Estimating reaction rates and  uncertainties for primordial 
nucleosynthesis}
\author{Kenneth M. Nollett$^{1,2}$ and Scott Burles$^2$}
\address{$^1$Department of Physics\\
 Enrico Fermi Institute, The University 
of Chicago, Chicago, IL~~60637-1433\\
$^2$Department of Astronomy and Astrophysics\\
Enrico Fermi Institute, The University of Chicago, Chicago IL 60637-1433}
\date{\today}
\maketitle

\begin{abstract}
We present a Monte Carlo method for direct incorporation of nuclear
inputs in primordial nucleosynthesis calculations.  This method is
intended to remedy shortcomings of current error estimation, by
eliminating intermediate data evaluations and working directly with
experimental data, allowing error estimation based solely on published
experimental uncertainties.  This technique also allows simple
incorporation of new data and reduction of errors with the
introduction of more precise data.  Application of our method
indicates that previous error estimates on the calculated abundances
were too large by as much as a factor of three.  Since uncertainties
in the BBN calculation currently dominate inferences drawn from
light-element abundances, the re-estimated errors have important
consequences for cosmic baryon density, neutrino physics, and lithium
depletion in halo stars.  Our direct method allows detailed discussion
of the status of the nuclear inputs, by identifying clearly the places
where improved cross section measurements would be most useful.
\end{abstract}

\pacs{26.35.+c, 98.80.Ft}
%\vskip 1in
]

%\twocolumn

\section{introduction}

Big-bang nucleosynthesis (BBN) is an important component of the hot
big-bang cosmology.  It provides a direct probe of events less than
one second after the big bang, as well as key evidence for the
existence of non-baryonic dark matter.  The success of big-bang
nucleosynthesis theory is indicated by the narrow range of cosmic
baryon density over which the observed abundances of the light
isotopes, D, $^3$He, $^4$He, and $^7$Li agree with their calculated
abundances.  This narrow range in the theory's single free parameter
(once the input physics is specified) was summarized in 1995, in units
of critical density, as $\Omega_Bh^2=.009$--.020\cite{cst}.  ($h$ is
the Hubble constant in units of 100 km/s/Mpc; this limit is also
customarily quoted in terms of the baryon-to-photon number ratio,
$2.5\times 10^{-10} < \eta < 6\times 10^{-10}$.)

The situation has changed dramatically since that time, with the
arrival of more precise astronomical measurements of D, He, and Li
abundances.  Of particular note are the precise measurements of the
deuterium abundances in high-redshift quasar absorption systems
\cite{burlestytler98a,burlestytler98b}.  While most of the deuterium
in the solar neighborhood has been subject to destruction in
pre-main-sequence stars, the composition of these objects is believed
to be nearly primordial.  (Deuterium is so weakly bound that BBN is
the only realistic site for cosmic production \cite{reevesetal}.)
Because the amount of deuterium produced in BBN depends strongly on
baryon density, these measurements allow a tight constraint on that
parameter --- presently at a level of 8\%, based on the measurements
of Burles and Tytler \cite{burlestytler98a,burlestytler98b} and the
standard estimation of theoretical errors.  In a few years, the
deuterium-inferred baryon density will be subject to comparison with a
similarly precise inference of the baryon density from observations of
the cosmic background radiation \cite{st98}.  This is an important
comparison, because the physical bases of these two inferences are
completely independent.

\begin{figure}
\centerline{\epsfig{file=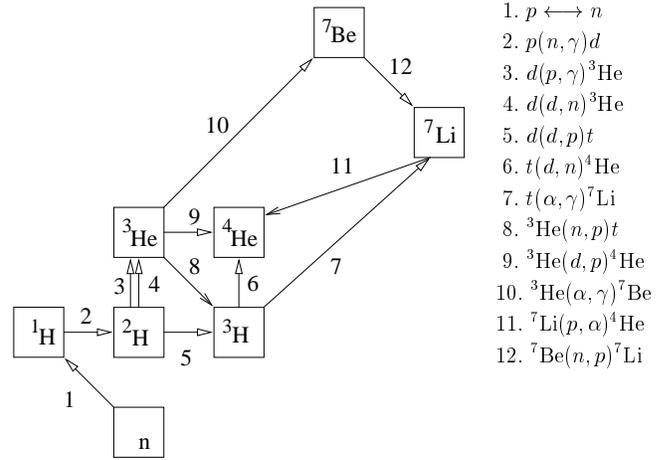,height=8.6cm,angle=270}}
\caption{The reaction network that determines yields in standard
BBN.}
\label{fig:network}
\end{figure}

The new state of affairs (with corresponding advances in He and Li
observations) has been described as a ``precision era'' for big-bang
nucleosynthesis \cite{st98}.  However, as the observational
uncertainties shrink, the uncertainties on the calculated abundances
begin to dominate.  Although the nucleosynthesis calculation itself is
straightforward, and has been well-understood for over three decades
now \cite{wagonerfowlerhoyle,wagoner73,kawano,kawano2,kr,skm},
``theoretical'' uncertainties arise from its nuclear cross section
inputs.  These inputs consist of cross sections (Fig. 1) which have
been measured in nuclear laboratories since the 1930's, with accurate
enough data for nucleosynthesis work dating mostly from the 1950's and
1960's.  It is well-known that stellar nucleosynthesis requires cross
sections below the Coulomb barrier, which cannot be directly probed in
the laboratory because the cross sections are too small.  In contrast,
BBN occurs at sufficiently high temperatures ($T\sim 10^9$ K) that
this is not a serious problem, and data are generally present at
exactly the energies where they are needed for a BBN abundance
calculation without any recourse to theoretical modeling.  In fact,
there is a sufficiently large body of precise data in the energy range
needed for BBN that the results of the calculation have remained
nearly the same since the standard code was first run by Wagoner in
1967 \cite{wagonerfowlerhoyle}, despite a slow trickle of new cross
section measurements.  Nonetheless, some uncertainty remains in the
calculations, arising from experimental uncertainties in determining
these cross sections.  These uncertainties range from about 5\% to
25\% in the cross sections, and propagate to errors of up to nearly a
factor of two (from lower to upper $2\sigma$ limits) in the case of
the calculated $^7$Li abundance.  (This is true of ``standard BBN,''
which contains nothing beyond the standard model of particle physics.
There are, of course, greater uncertainties if alternative -- {\it
e.g.} baryon-inhomogeneous -- scenarios are considered
\cite{jedamzik95,kurki97}.)

The present ``industry-standard'' error estimation for the inputs was
done by Smith, Kawano, and Malaney (hereafter SKM, Ref. \cite{skm}) in
1993, using the Monte Carlo error propagation method applied earlier
by Krauss and Romanelli \cite{kr}.  This was a landmark work because
it examined all the nuclear inputs critically and in detail, and it
placed quantitative error estimates on all the inputs.  There has been
a small number of new cross section measurements since
\cite{tag-brune,dpg-schmid,dpg-ma,he3np-brune}, which have been
unevenly incorporated in subsequent work.  Subsequent authors have
continued to use the SKM rates and errors for most or all reactions.
A few have substituted the results of a single new measurement in
place of the corresponding SKM evaluation of all experiments for the
given reaction.

Because the SKM uncertainties are used so widely to draw quantitative
conclusions concerning many aspects of cosmology and particle physics,
it is important to examine their assumptions and attempt improvement
as well as to maintain some up-to-date set of uncertainties.  The SKM
work proceeds as follows: Cross section data for the key reactions
(Fig.~\ref{fig:network}) are gathered from an
extensive survey of the literature.  These data are fitted to
standard, if very approximate, theoretical expectations concerning
their energy dependence (typically a low-order polynomial, for the
low-energy $S$-factor), which has been widely used in previous work
\cite{fcz,fcz2,fcz3,cf}.  Some arbitrary choices are made concerning
inclusion and exclusion of data points in the fits, with a view to
making these fits accurate over the energy range critical for BBN.
Error estimation begins with formal error estimates on the fitted
parameters, but is modified where necessary to include most or all
data points within two-sigma error curves.  This conservative approach
justifies some arbitrariness in weighting the data sets.  The SKM
errors on the cross sections are summarized (with two exceptions) as
uncertainties in the overall normalization, and then propagated
through the basic BBN calculation by a Monte Carlo procedure with
normalization values drawn from Gaussian distributions.

We begin by listing principles for a better treatment of nuclear data
in a BBN calculation, commensurate with the goals of ensuring that the
confidence limits being quoted in cosmological work are meaningful,
and of making a direct link to the nuclear measurements.  Because the
energy dependences of the cross sections are not in question in most
cases, our main goal is to arrive at a suitable method of estimating
errors in the BBN yields.  The desirable qualities for such a method
are are: 1) Nuclear data, and their published uncertainties, should be
incorporated into the BBN calculation as directly as possible, so that
errors in the calculation are directly linked to the nuclear data set,
and the results of the BBN calculation reflect as few arbitrary
choices as possible; 2) minimal assumptions concerning functional
forms of cross sections (as functions of energy) should be used, where
there are enough data to characterize these functional forms (almost
all cases); 3) there should be accounting for correlated errors in the
data, since they are ubiquitous; 4) the incorporation of future data
with smaller formal errors into the inputs should reduce the
uncertainty estimates in the calculation; and 5) the prescription
should be no more conservative than necessary, given the precision
with which abundances are now being measured.  The procedure of SKM,
outlined above, does reasonably well on point number 2.  Point number
5 is something of a matter of taste (although there is a mismatch in
levels of conservatism between SKM and the errors quoted by
astronomers).  However, the SKM procedure does not answer well to our
other requirements.  The subjective assignment of errors causes
trouble on points 1, 3, and 4.  In particular, the conservative SKM
method is not re-applicable in a way that narrows error estimates as
new data are incorporated in the data set, unless old data are thrown
out, and so it does not satisfy point number 4.  Poor performance on
this point discourages the production of more precise measurements,
because it is unclear how new data will affect error estimates on
reaction rates.

To improve on the SKM work, we propose a new prescription for treating
the nuclear data that uses the measured cross sections directly in the
standard big-bang nucleosynthesis calculation.  It is a Monte Carlo
technique that uses ``realizations'' of the full nuclear data set for
the key reactions identified by SKM and Krauss and Romanelli to derive
reaction rates.  We extract final nucleosynthesis yields and
confidence limits from distributions of yields corresponding to the
distributions of cross sections implied in quoted experimental errors.
While there may be some question of uniqueness, our method does
provide a useful and unambiguous prescription for linking the results
of the BBN calculation to its nuclear inputs.  This method is very
much in accord with the stated goals of earlier work, both of SKM and
of Krauss and Romanelli \cite{kr}, and it reflects increasing computer
speed.  In the words of Krauss and Romanelli, ``unless error estimates
are tied to a direct analysis of the data one cannot accurately gauge
the margin for improvement.''

The rest of this paper is organized as follows: In Section
\ref{sec:method}, we describe our Monte Carlo method for computing
light-element yields and confidence intervals from the nuclear data.
In Section \ref{sec:data}, we discuss the nuclear inputs in detail,
emphasizing differences from the analysis of SKM.  In Section
\ref{sec:results}, we discuss the results of a standard BBN
calculation using our method.  In Section \ref{sec:conclusions}, we
discuss the implications of our work both for the BBN calculation and
for cosmology in general.  

\section{method}
\label{sec:method}

For our purposes, the nuclear inputs for a BBN calculation come in the
form of angle-integrated cross sections, $\sigma(E)$, where $E$ is
reaction energy, reported in the published literature.  The quantities
needed to evolve abundances in a reaction network are
thermally-averaged cross sections,
\begin{equation}
\langle\sigma v\rangle = \sqrt{8 \over {\pi \mu (kT)^3}} \int
 \sigma(E) \, E \, e^{-E/kT} dE,
\end{equation}
where $\mu$ is the reduced mass and $k$ is Boltzmann's constant.  

In the case of charged particles, the procedure typically followed in
the past to obtain these thermally-averaged rates has been as follows:
One first re-parameterizes the measured cross sections as $S$-factors,
\begin{equation}
S(E)=E\sigma(E) e^{2\pi\zeta},
\end{equation}
where $\zeta=Z_1Z_2\alpha/v$, to remove the strong Coulomb dependence
of the cross section.  Here, $\alpha$ is the fine structure constant,
$Z_i$ are atomic numbers of the nuclei, and $v$ is their relative
velocity.  The resulting function is then fitted to a low-order
polynomial plus resonant terms to obtain a form which can be
integrated.  (This functional form has sometimes been used for
extrapolation to low energy, although that is discouraged.)  The
integration of the nonresonant rate is then performed using a
saddle-point approximation with lowest-order corrections
\cite{fcz,fcz2}, or else performed numerically and fitted to the
functional form one gets from the saddle-point integration.  (The same
considerations apply for neutron-induced cross sections, with the
exception that they are fitted by a low-order polynomial in velocity
before integration, and the integrations are then exact.)  Resonant
cross sections are fitted to Breit-Wigner or single-level $R$-matrix
forms, integrated numerically, and fitted to analytic forms.  While these
functional forms are useful for disseminating evaluated reaction rates
in printed form, and have therefore become somewhat standard, they are
not ideal for producing rates valid over a wide range of temperatures
or for arbitrary $S(E)$.

Both SKM and Krauss and Romanelli used this sort of procedure,
including most of the extant data in the fits, and being careful that
the fits were performed over the energy range needed for BBN
calculations.  They followed it up by estimating uncertainties on the
reaction rates, expressed as errors in overall normalization (except
for two cases of energy-dependent errors in SKM), which are
appropriate to the BBN energy range.  They then estimated the
corresponding uncertainties in BBN yields by varying the rates
according to these error estimates in Monte Carlo BBN yield
calculations and examining the distribution of output abundances.  The
Monte Carlo approach was originally deemed necessary because it is not
clear without doing the calculation whether the errors combine
linearly.  The recent work of Fiorentini {\it et al.}  \cite{sarkar}
indicates that in fact normalization errors on the reaction rates can
be propagated linearly through the BBN calculation, and the results
are very close to the corresponding Monte Carlo results.

Our method differs from that of these earlier efforts in several ways:
we fold the process of characterizing reaction rates into the same
Monte Carlo process that calculates abundances, we try to make the
least possible number of assumptions about the functional forms of
cross-section energy dependences, and we try to keep error estimation
based strictly on quoted errors in the nuclear data set, including
correlated errors explicitly.

Our Monte Carlo calculations sampled the space of nuclear cross
sections indicated by the nuclear data, and the result was a
distribution of output abundances, from which we derived confidence
limits.  Each calculation sampled 25,000 points in this space.  (This
was a number that gave us smooth 95\% cl curves as a function of
baryon density.)  A single Monte Carlo sample consisted of the
following steps:

\begin{enumerate}

\item For every measured cross section of every reaction (every
$\sigma(E)$ in our database), a random number was drawn from a
Gaussian distribution whose mean is the reported cross section at that
energy, and whose variance is the reported variance for that point.
Also, for each data set (collections of $\sigma(E)$ values from a
given experiment), a random number was drawn from a Gaussian
distribution whose variance was the normalization error shared by
those points, and all cross sections in that data set were multiplied
by this random normalization.  This provided accounting for correlated
errors.  The database of cross sections and uncertainties is described
in Sec. \ref{sec:data} below.

\item After ``synthetic'' cross section data were chosen, smooth
representations of these data were created for integration.
Specifically, the $S(E)$ (or $\sigma v$, for neutron-induced
processes) curve for each reaction was fitted to a piecewise
polynomial (in $B$-spline representation) as a function of energy.
The spline broke up the energy axis into several (typically less than
ten) segments, generally evenly-spaced in $\log E$, assumed a
polynomial of order 3--5 within each segment, and forced continuity of
derivatives across the segment boundaries.  This curve was fitted to
the simulated data by the customary weighted linear least-squares
technique.  Note that this approach is ``theory-free'' in the sense
that the only assumption we have made about the cross sections is that
they are sufficiently smooth functions of energy to be represented by
the chosen piecewise polynomial.  The $S$ factor re-parameterization is only a
re-parameterization, and is converted back to cross sections after
splines are fitted.  Although the number of variables for our smooth
representation was chosen by eye for each reaction individually, we
expect that small variations in the functional form of the $S$ factor
which depend on choices made in the fitting are smoothed out by the
subsequent thermal averaging and Monte Carlo sampling.
 
\item The smooth representation of each cross section was integrated
numerically at many different temperatures (ten per decade), and the
resulting reaction rates fitted to a smooth representation as a
function of temperature.  This allowed subsequent calculation of rates
by interpolation without expensive integrations.

\item Finally, these reaction rates were fed into a standard BBN code
\cite{kawano,kawano2}, which used them to calculate yields.  Note that
since independent integrations of different numbers are performed for
each Monte Carlo step, any (non-systematic) integration errors are
accounted for in the Monte Carlo process.

\end{enumerate}

When the code was finished, we extracted 95\% confidence limits from
the distribution of yields.  This process is similar to a treatment of
fitting errors described in {\it Numerical Recipes} \cite{nr}.

We treated three reactions somewhat differently from the rest.  First,
we used the most recent experimental value for the neutron lifetime,
$885.4\pm 2$ s, to derive values for the neutron-proton
interconversion rates, choosing Monte Carlo values for this rate
according to a Gaussian distribution with this mean and variance.
These rates affect only the final $^4$He abundance significantly.
This is also the only process $^4$He yields are sensitive to at likely
values of baryon density.  Lopez and Turner \cite{lopezturner98} have
incorporated in a coherent and consistent way a number of small but
important physics and numerical corrections to the weak rates, and we
take yields for this nuclide from their work.  Second, there was not
enough data coverage to use our technique for proton-neutron capture,
$p(n,\gamma)d$, so we used a theoretical model of this process as
described in Sec.  \ref{sec:png} below.  Finally, the apparent
presence of systematic discrepancies in measurements of the $^3{\rm
He}(\alpha,\gamma)^7{\rm Be}$ cross section required the special
treatment described in Sec. \ref{sec:he3ag}, after applying our
standard technique to a subset of the data.

\section{nuclear inputs}
\label{sec:data}

\subsection{The Database}

The nuclear data used in our work were obtained from a comprehensive
survey of the experimental literature from approximately 1945 onward.
Many of the numerical values were obtained from the on-line CSISRS
database \cite{csisrs}.  However, even in these cases, we incorporated
data sets only after reading the original sources carefully.  For
almost all reactions, the data sets included were the same ones found
in SKM.  There are three reactions, $d(p,\gamma)^3{\rm He}$, $^3{\rm
He}(n,p)^3{\rm H}$, and $t(\alpha,\gamma)^7{\rm Li}$, for which more
recent cross section data than those used by SKM exist.  Some
subsequent BBN calculations ({\it e.g.,} Ref. \cite{cst}) have
incorporated the latest measurement of $t(\alpha,\gamma)^7{\rm Li}$
\cite{tag-brune} by replacing the SKM fit with a reaction rate and
uncertainty based on that measurement alone.  These numbers have not
found their way into all subsequent work ({\it e.g.,}
Ref. \cite{sarkar}), and we know of no calculations that have
incorporated either of the recent TUNL measurements of the
$d(p,\gamma)^3{\rm He}$ \cite{dpg-schmid,dpg-ma} cross section (save
one reported in this last experimental reference).  One of our goals
is to provide a new standard calculation which incorporates these
measurements so that they are not ignored in future theoretical work.

Another of our major goals is to incorporate explicitly information
concerning known systematic errors in the cross section measurements,
which indicate correlations among data from a given experiment.  The
incorporation of this information into our Monte Carlo calculation is
described in Sec. \ref{sec:method} above.  Here, we describe the
origin of the numbers used.  Data sets which include measurements of
the cross section for the same process at several energies contain
both {\it shared} normalization errors (which affect all points in
that data set) and {\it unshared} point-to-point errors (which vary
from one point to the next, and arise primarily from counting
statistics).  Because it is generally much more difficult to determine
the absolute normalization of a cross section than it is to measure
its energy dependence, most of the uncertainty in a given cross
section is usually in its normalization.  It is therefore important to
treat normalization errors in the data properly when combining data
sets, especially if the data sets are of different sizes.  It is not
correct to apply the quoted ``total error'' for each point in a fit
when the correlated errors have been added in (the usual case for
published data).  Our calculation assumes Gaussian-distributed errors
as the simplest assumption in every case, although one might expect
that normalization errors, in particular, will not be
Gaussian-distributed.  Note that in other contexts, one typically
allows a floating overall normalization for each data set in fitting
nuclear data.  Since the energy dependences are well-determined for
almost all of the eleven key cross sections, we are only interested in
experiments which measure the cross section normalization.

Separate treatment of shared and unshared errors involves some
difficulties.  These include the fact that systematic effects may not
be well-quantified, and the fact that experimental data have often
been presented in ways incompatible with this goal.  Several of the
experimental sources (especially the more recent ones) explicitly
state shared normalization errors separately from unshared errors.  In
some cases, quoted normalization errors had to be subtracted from
quoted total errors, which are quadrature sums with unshared errors.
Other experimental sources did not explicitly add up the shared
errors, but they usually listed the percent contribution from each
major item in the error budget separately.  These could usually be
added up to provide normalization errors, with a small amount of
guesswork as to which contributions to place in each category.  For
example, detector efficiencies and target chemistry are often readily
identifiable as sources of shared error, in those cases where they
are.  Because errors are almost always added in quadrature, our
procedure was not strongly dependent on identifying which category
(shared or unshared) each individual contribution should be placed in,
so long as we were careful that estimates of unshared errors did not
become too small.  We are confident that the sizes of errors adopted
for our database are all roughly correct.  Our method required that we
exclude a small number of data sets which did not provide enough
information for such a breakdown of error sources (but we were biased
toward keeping data sets, especially where there were few measurements
of a cross section).  We also excluded data sets which only represent
measurements of relative cross sections.  Fortunately, we did not need
to exclude a large number of data sets for any reason.

We have not incorporated any of the substantial body of theoretical
knowledge (ranging from unitarity constraints to ``microscopic''
reaction models) which exists for some of these processes into the
database, except for the process $p(n,\gamma)d$.  The reason for this
is that there is little need for theoretical evaluations where we
already have large amounts of data, especially since errors are often
hard to assign to models.  A typical use of theoretical models is to
calculate an energy dependence for a cross section, test it by
comparison with relative cross section measurements, and then
normalize it by absolute cross section measurements.  This is more
constraining than our approach, since it allows data taken at all
energy ranges (and sometimes in other reaction and scattering
channels) to affect the evaluated cross section at a given energy, but
it would require a significantly larger and less straightforward
effort.  In this regard, our ``theory-free'' approach may still
provide conservative error estimates for some reactions.

Because of all these concerns, particularly those concerning unknown
systematics and arbitrariness of error assignment in experiments,
there can probably be no unique evaluation of rates and their errors.
We have attempted to make estimates with the desirable qualities
listed above, and to improve on previous efforts.

\subsection{Reaction Sensitivities}

Because our approach to the BBN calculation is very closely tied to
the data, we are able to make very precise statements about exactly
which inputs the calculated yields are sensitive to (at least for the
``standard BBN'' calculation we have done).  In particular, we have
quantified this sensitivity with what we call ``sensitivity
functions'' for each reaction and nuclide.  These may be regarded as
functional derivatives of BBN yields with respect to reaction $S$
factors.  For a given reaction and nuclide, we calculate one of these
functions numerically by adding a small amount to the reaction $S$
factor over a narrow bin in energy, and computing the resulting
primordial abundance of that nuclide.  The sensitivity function is the
fractional difference between this yield and the unperturbed yield, as
a function of the location of the energy bin.  In the limit of very
narrow energy bins, this is a functional derivative.  The functions
indicate quantitatively the exact energies at which each process is
important in standard BBN, and therefore the exact energies at which
precise cross section measurements are needed to produce precise BBN
calculations.  We denote the sensitivity functions by $g_2(E)$ for D/H
and $g_7(E)$ for $^7$Li/H, and they are shown for all eleven cross
sections below.

The general behavior of these functions is clear: Above some
temperature in the vicinity of $10^9$ K, the actual reaction rates do
not matter because the reactions are in thermal equilibrium, with
forward and reverse reactions proceeding at the same rate.  When the
density and temperature drop to some point at which a particular
reaction falls out of equilibrium, its rate may become a determinant
of the nuclide abundances.  Finally, there comes a point when
temperatures and densities are too low for a reaction to change
abundances at all.  The approximate effective energies of reactions
for nuclear burning at a given temperature correspond to the customary
``Gamow peak'' of nuclear astrophysics \cite{fcz}, and what we see in
the sensitivity function is often the Gamow peak, convolved with
the distribution of temperatures at which a given reaction is active,
but out of equilibrium.  The rate sensitivities are also functions of
the baryon density, as indicated in the plots of
Sec. \ref{sec:reactions}.

\subsection{Statistical tests}
\label{sec:chi2}

Our new method for calculating BBN abundances should be examined to
test the robustness of its results and the consistency of the input
nuclear data.  We therefore applied it to fake data drawn from the
assumed source distribution of the actual data.  We generated (for
each reaction) fake data that were Gaussian-distributed about our
highest-probability curve.  Correlated data were multiplied by
appropriate Gaussian-distributed common normalizations.  The fake data
were placed at the same energies as the actual data, and the
distributions were based on the quoted errors of the corresponding
actual data.  For {\it each collection} of fake data, we applied our
Monte Carlo curve-fitting method, and computed the means, $\mu(b_i)$,
and standard deviations, $\sigma(b_i)$, of the $B$-spline coefficients
describing the fitted curves.  ($b_i$ denotes the coefficient of the
$i$th basis function.  Note that the local support of $B$-spline basis
functions \cite{deboor} makes each of these coefficients a sort of
weighted local average of the fitted function, so that $\mu(b_i)$ and
$\sigma(b_i)$ reflect the means and standard deviations of the fitted
curves directly.)  After performing this procedure for many ($\geq
1000$) collections of fake data, we computed the means,
$\overline{\mu(b_i)}$, $\overline{\sigma(b_i)}$ and standard
deviations $\Sigma(\mu(b_i))$, $\Sigma(\sigma(b_i))$ {\it over the
fake-data distribution} of the $B$-spline coefficient means $\mu(b_i)$
and standard deviations $\sigma(b_i)$ derived from each choice of fake
data.

Comparison of these numbers with the means and variances of $B$-spline
coefficients generated as intermediate numbers by our BBN code
indicates that the means reproduce the assumed curves consistently
with our error estimates in all but one case.  The mean variances
$\overline{\sigma(b_i)}$ of the $B$-spline coefficients generally
agree with the standard deviations of $B$-spline coefficients from our
BBN procedure (which are $\sigma_i$ for the actual data); where they
differ, our BBN procedure gives slightly larger standard deviations.
The standard deviations of the means, $\Sigma(\mu(b_i))$, were
virtually always identical to the mean standard deviations,
$\overline{\sigma(b_i)}$.  In other words, the distributions of
$S$-factor curves from our procedure do not change drastically when
the experimental data are drawn from other points in the same
distributions from which we assume the actual data to be drawn.
Serious errors in reproducing the assumed curves occurred only for
$^3{\rm He}(d,p)^4{\rm He}$.

We repeated the same numerical experiment, this time simulating a
factor-of-two underestimation of all normalization errors by adding
extra scatter to the fake data (doubling all of the normalization
errors).  In applying our Monte Carlo method to these fake data, we
assumed the quoted normalization errors, not the inflated ones.  The
most noticeable results were an increase in the standard deviations of
means $\Sigma(\mu(b_i))$ by a factor of about $\sqrt{2}$, and slighly
poorer reconstruction of several coefficients for
$t(\alpha,\gamma)^7{\rm Li}$.  Very little else changed.  The standard
deviations over fake data of the $B$-spline variances,
$\Sigma(\sigma(b_i))$, also increased in some cases, but were seldom
much more than 10\% of $\overline{\sigma(b_i)}$.  In other words, the
variances of the $S$-factor curves from our BBN procedure are
sensitive mainly to the error estimates rather than the quoted cross
sections, while the mean curves are less sensitive to the error
estimates.  Underestimated errors result in $S$-factor curves that
vary less than they should in our BBN calculation, but probably never
by factors of more than 1.5.

Although the results above suggest that our method is not overly
sensitive to unexpected scatter in the data, we still wish to address
the question of consistency of the nuclear data.  We calculate a
chi-squared statistic with the data and the ``best-fit'' model curve
describing the un-altered data.  This is not a goodness-of-fit test
for our modelling, because the individual curve plays no important
role in our calculation.  It is, rather, a benchmark by which to
examine the consistency of data measured at different energies.
Initially, we calculated this chi-squared statistic using the full
non-diagonal error matrix, as described in Ref. \cite{chi2c}.
However, as this reference shows, chi-squared calculations using the
non-diagonal error matrix are not appropriate when the correlations
are in the form of shared normalizations.  Fits minimizing this
statistic are almost always lower than the data when there are large
normalization errors.

The statistic that we did apply is the ordinary $\chi^2$ statistic,
defined as
\begin{equation}
\chi^2=\sum_i \frac{(S(E_i)-S_{\rm model}(E_i))^2}{\sigma_i^2},
\label{eq:chi2}
\end{equation}
where $S(E_i)$ are the data for a given reaction, measured at energies
$E_i$, $S_{\rm model}(E)$ is the best-fit model curve, $\sigma_i$ is
the quadrature sum of normalization and point-to-point errors for each
datum, and the sum is over all data for a given reaction.  Because of
the correlated errors, we do not expect this statistic to be
distributed as a formal chi-squared distribution.  We therefore
constructed the expected distribution by varying fake data about the
best-fit curve according to the quoted uncertainties as above.  This
allowed us to assess consistency of the data by determining the
likelihood of the actual value of $\chi^2$, given our assumptions
about the distribution from which the actual data were drawn.  The
results of computing $\chi^2$ for each reaction $S$-factor and
comparing it to the fake data distribution are shown in Table
\ref{tab:fits}.  From these results, we conclude that the only clear
case of trouble is $^3{\rm He}(d,p)^4{\rm He}$.  Note that this cross
section has no noticeble effect on the calculated deuterium abundance,
and very little effect on the calculated $^7$Li abundance.  It is more
important for $^3$He, which is not as yet amenable to high-precision
treatment as a product of BBN.

The reader may object that we do not have a procedure to inflate the
errors where the $\chi^2$ test indicates that they may have been
underestimated, or to discard data sets which contribute
disproportionately to $\chi^2$.  We have resisted adopting such
procedures for several reasons.  The first is that it would be
difficult to arrive at a unique and meaningful prescription to
determine by how much to inflate errors, or which data to discard.
(The customary multiplication by the square root of the $\chi^2$ per
degree of freedom is not suitable because $\chi^2$ is not distributed
as a formal $\chi^2$ distribution, and because the factor should in
principle be a function of energy.)  The second reason is that we
would like to leave any systematic problems with the existing data,
and their properties with regard to generating formal errors for BBN,
as distinct issues.  A database which performs well on both counts is
desirable for producing reliable BBN calculations.  On the other hand,
discarding data is something we will not do without good reasons from
the nuclear physics literature, even though disproportionately large
fractions of the $\chi^2$ statistics usually do come from individual
data sets in our database.  Finally, there is only one ``two-sigma''
unlikely $\chi^2$ in Table \ref{tab:fits}, although there are more
``one-sigma'' unlikely values of $\chi^2$ than expected.
%\widetext
\begin{table*}
\caption{For each reaction, the number of data points, characteristics
of the piecewise polynomial descriptions, values of $\chi^2$, and
probabilities that $\chi^2$ exceeds its measured values (based on
20,000 Monte Carlo samples).}
\label{tab:fits}
\begin{tabular}{ccccdd}
 & & {Spline} & & & {Probability}\\
Reaction & Points& segments & order & $\chi^2$  &   of exceeding $\chi^2$ \\
\tableline
p(n,$\gamma$)d & --- & 1 & 10 & \multicolumn{2}{c}{(not fitted to lab data)} \\
d(p,$\gamma$)$^3$He\tablenotemark[1] & 16 & 5 &3&3.40 & 93\% \\ 
d(d,p)$^3$H & 134 & 6 & 3   & 108.0  & 68.7\%  \\ 
d(d,n)$^3$He & 130 & 6 & 3 & 106.7 &64\%   \\ 
$^3$He($\alpha$,$\gamma$)$^7$Be\tablenotemark[2] 
& 118 & 4 & 3 & 138.3 & 37\% \\ 
$^3$He(d,p)$^4$He & 111 & 9 & 4 & 315.0 & 0.5\%  \\ 
$^3$He(n,p)$^3$H & 240 & 5 & 5  & 106.0 & 91\%  \\ 
$^7$Li(p,$\alpha$)$^4$He & 105 & 6 & 3 & 193.1 & 11\% \\
$^7$Li(p,n)$^7$Be & 137 & 29 & 3 & 223.0 & 16\% \\ 
$^3$H($\alpha$,$\gamma$)$^7$Li & 55 & 6 & 3 & 129.5  & 12\%  \\ 
$^3$H(d,n)$^4$He & 213 & 15 & 3 & 292.8  & 13\%  \\ 
\end{tabular}
\tablenotetext[1]{This reaction is a special case, since none of the
experiments in our database overlap in energy.}
\tablenotetext[2]{Only capture photon data are described here.}
\end{table*}

\subsection{Individual Reactions}
\label{sec:reactions}

The following section is dedicated to detailed discussions of the data
sets for each individual reaction.  For convenience, we have followed
the same ordering as SKM in discussing individual reactions.  Because
the available data have changed very little since the work of SKM, we
emphasize differences from their analysis and new insights provided by
our examination of the inputs.  We also list the references from which
we compiled our database.  Unless explicitly discussed, any omission
was because either a normalization error could not be extracted from
the original source, or the experiment measured only relative cross
sections (or, trivially, the experiment contained no data in the
relevant energy range).

\begin{figure}
\centerline{\epsfig{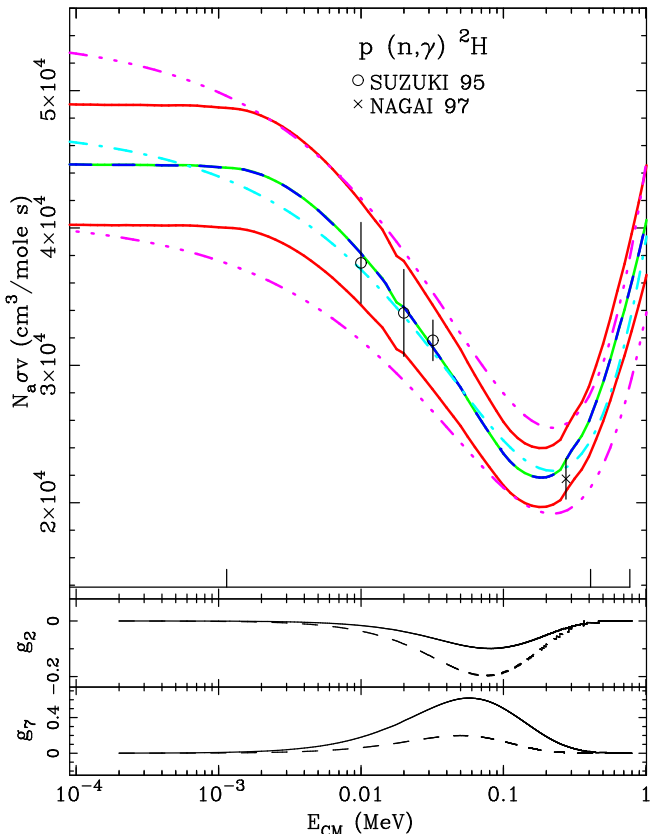}}
\caption{Theoretical curve from Hale {\it et al.} \protect\cite{png},
SKM fit to that curve with $2\sigma$ errors, and current experimental
data in the BBN energy range \protect\cite{png-suzuki,png-nagai}.  An
explanation of our nuclear data figures may be found near the
beginning of Sec.  \protect\ref{sec:reactions}.}
\label{fig:png}
\end{figure}

For each reaction, we show in a graph
(Figs. \ref{fig:png}--\ref{fig:li7pa}) the input data, the
``best-fit'' curves and 95\% confidence limits inferred from our Monte
Carlo procedure (solid curves), and (where available) the
corresponding curves from the SKM analysis (dash-dotted curves), and
curves from the ENDF-B/VI \cite{endf} evaluation (dashed curves).  The
tick marks below the data in each graph show the region over which the
integral must be performed to get the yields correct to one-tenth of
the total uncertainty in all abundances (inner tick marks), and to get
the yields correct to one part in $10^5$ (outer tick marks).

In the lower panels of each graph, we show the sensitivity functions
at baryon densities of $\Omega_Bh^2=0.019$ (solid curves) and
$\Omega_Bh^2=0.009$ (dashed curves) for D and $^7$Li.  They represent
very well the contribution of the $S$ factor uncertainties to the
final yield uncertainties as functions of reaction energy in the sense
that, to a good approximation, the convolutions of these curves with
the 95\% cross section limits gives the 95\% yield limits due to that
reaction indicated by the Monte Carlo study described below.  The $g$
functions shown in the plots have been multiplied by energy in MeV so
that relative areas under the curves can be accurately judged on the
logarithmic scale on which we have plotted the nuclear data.  Some
references contained two or more distinct data sets with different
normalization errors; in these cases, an extra number after
publication year distinguishes symbols from the same publication.  All
energies are in the center-of-mass frame.

\subsubsection{$p(n,\gamma) d$}

\label{sec:png}

As noted above, this reaction required special treatment.  It could
not be subjected to our piecewise splining or Monte Carlo sampling of
data points because of an extreme scarcity of data in the $E_{CM} <
300$ keV range.  This is despite recent efforts to measure this
crucial cross section \cite{png-suzuki,png-nagai} in the relevant
energy range, which have so far resulted in only four data points.  We
have chosen as the cross section for this process its evaluation in
the ENDF-B/VI database \cite{png}.  This cross section derives from a
theoretical model computed sometime around 1970, constrained by $np$
scattering phase shifts, along with a few photodisintegration data
(known to have systematic problems), and it has seen a minor update to
match the current value of the well-measured thermal-neutron capture
cross section \cite{png-mughabghab}.  No documentation for this model
survives.  The crucial energy range for BBN corresponds to a
changeover in reaction mechanisms (from $M1$ capture to $E1$ capture)
for this process, so one might expect the validity of the evaluation
to be most in question there.  It has held up remarkably well in light
of the recent measurements, but its authors view this as a fortuitous
coincidence \cite{hale}.  SKM estimated a 5\% uncertainty on this
evaluation, based on uncertainties quoted in earlier tabulations of
the cross section for practical use.  It is very difficult to trace
the origin of this number, but we have adopted a 5\%
Gaussian-distributed normalization error as the uncertainty in this
cross section --- both for consistency with SKM, and because it is not
too far from an estimate by the evaluation's authors of ``at least
10\%'' \cite{hale}.  (Note that the 7\% total uncertainty quoted by
SKM includes errors in further fitting and integration.  We do no
further fitting, and we estimate an error of less than 1\% in our
numerical rate integrations.)

\subsubsection{$ d(p,\gamma)^3{\rm He}$}

\begin{figure}
\centerline{\epsfig{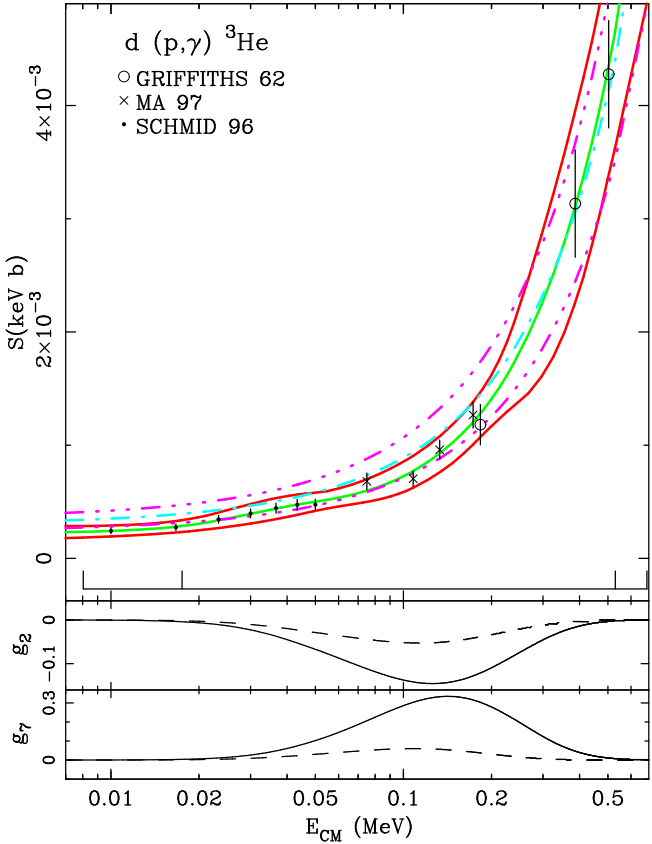}}
\caption{$S$ factor data and fits for $d(p,\gamma)^3{\rm He}$.}
\label{fig:dpg}
\end{figure}

The data set for this reaction is not well suited to our method.  Of
the three experiments performed before 1990, the experiment of Bailey
{\it et al.} \cite{dpg-bailey} is only a relative measurement, and not
suited to our method; the low-energy experiment of Griffiths {\it et
al.}  \cite{dpg-griffiths-low} deconvolved thick-target data with a
simple model of combined $S$- and $P$-wave capture at a sharp nuclear
surface.  The resulting energy dependence is in conflict with both the
more recent measurements of Schmid {\it et al.}  \cite{dpg-schmid} and
the three-body microscopic calculation of Viviani {\it et al.}
\cite{viviani}.  The recent TUNL experiments \cite{dpg-schmid,dpg-ma}
also suggest that these earlier thick ice target experiments used the
wrong stopping powers, resulting in cross sections about 15\% too
high.  We excluded the Griffiths {\it et al.}  and Bailey {\it et al.}
data sets for these reasons, but only after checking that their
omission did not alter our results drastically.  As indicated below,
this reaction contributes a large portion of our uncertainty
estimates.  The only other experiment in our database for this
reaction is from Ref. \cite{dpg-griffiths-high}.

The sparseness of the data may be a cause for concern regarding our
treatment of this cross section.  However, it is encouraging that our
splines indicate much the same energy dependence as the microscopic
calculation \cite{viviani,dpg-ma}, so the sparseness is probably not
such a big problem.  Our errors reflect the $\sim 8\%$ normalization
errors in the best measurements of the cross section.

\subsubsection{$d(d,n)^3{\rm He}$}

\begin{figure}
\centerline{\epsfig{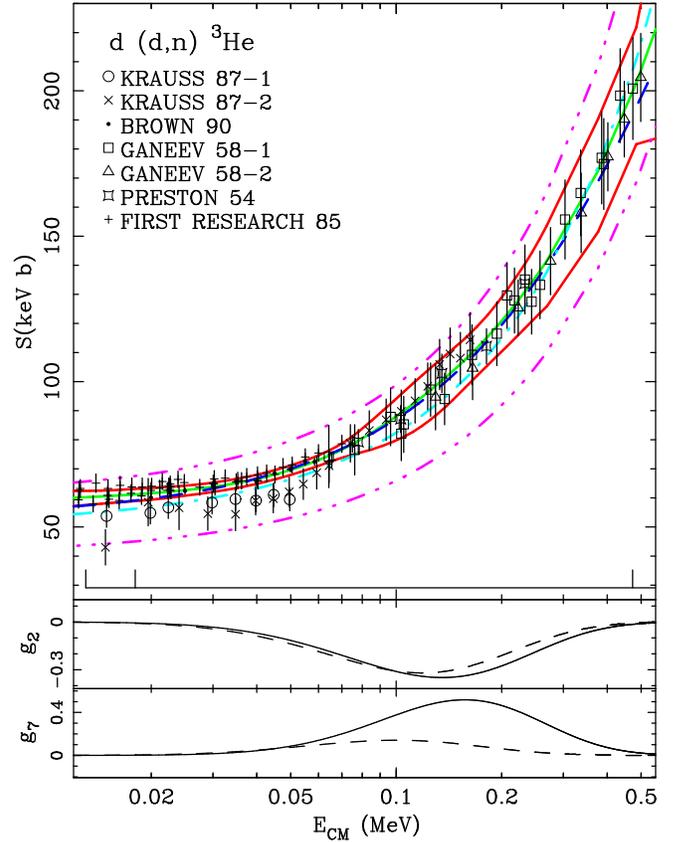}}
\caption{$S$ factor data and fits for $d(d,n)^3{\rm He}$.}
\label{fig:ddn}
\end{figure}

The deuteron-deuteron reactions have been measured very extensively
for fusion applications, and they are especially well-constrained
below about 60 keV by the precise measurements of Brown and Jarmie
\cite{dd-brownjarmie-prc}.  The SKM errors were based on an apparent
discrepancy between these data and those of Krauss {\it et al.}
\cite{dd-krauss}.  Brown and Jarmie make a similar assessment of the
situation, and recommend renormalizing the Krauss {\it et al.}  data
to their more precise data (which we do not do).  It is important to
note that most of the error in both cases is contained in
normalizations, so the case is not a 10\% discrepancy between ten data
points with $8\%$ errors and eleven points with 1.5\% errors.  Note
that the sensitivity functions peak at 100 keV or above, beyond the
Brown and Jarmie data.  With the exception of Krauss {\it et al.}
(with an $ 8\%$ claimed normalization error), all of the data in this
region date from before 1960, and their large errors and scatter are
responsible for the large contribution of this reaction to the
uncertainties in BBN yields.  Other data for this reaction are from
Refs.  \cite{dd-ganeev,dd-preston,dd-first}.

\subsubsection{$d(d,p)t$}

\begin{figure}
\centerline{\epsfig{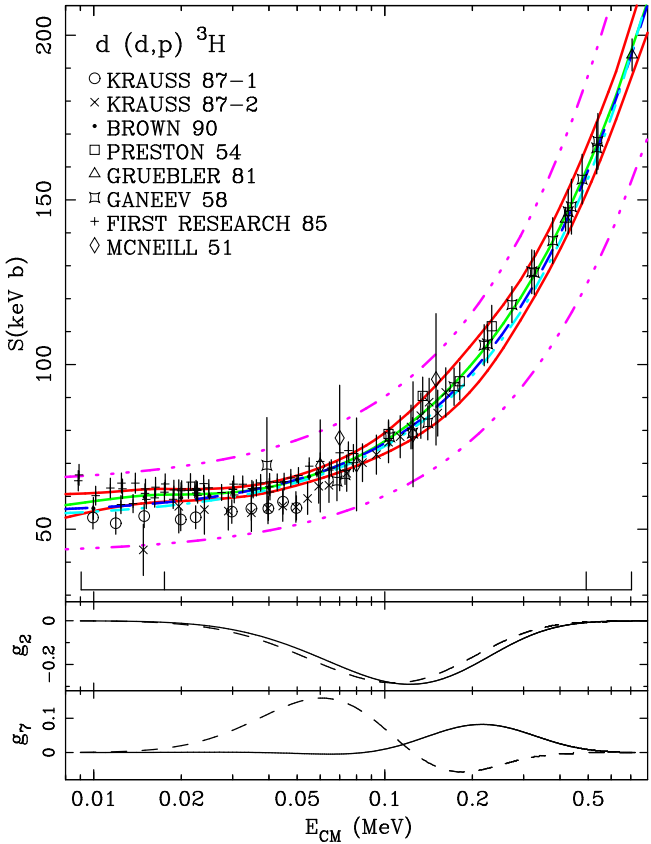}}
\caption{$S$ factor data and fits for $d(d,p)t$.}
\label{fig:ddp}
\end{figure}

This cross section is generally measured concurrently with that of
$d(d,n)^3{\rm He}$, so most of the discussion for that reaction
carries over to this one.  The sensitivity functions here are more
complicated, but there is still a large contribution at energies
greater than 100 keV.  There are also fewer data above 100 keV in this
case, reflecting the use of neutron-specific detection methods in some
$d(d,n)^3{\rm He}$ experiments.  Again, few of the data which are
present in this range come from modern experiments.  The data for this
reaction were taken from Refs.
\cite{dd-brownjarmie-prc,dd-krauss,dd-ganeev,dd-preston,dd-first,dd-gruebler,dd-mcneill}.

\subsubsection{$^3{\rm He}(n,p) t$}

\begin{figure}
\centerline{\epsfig{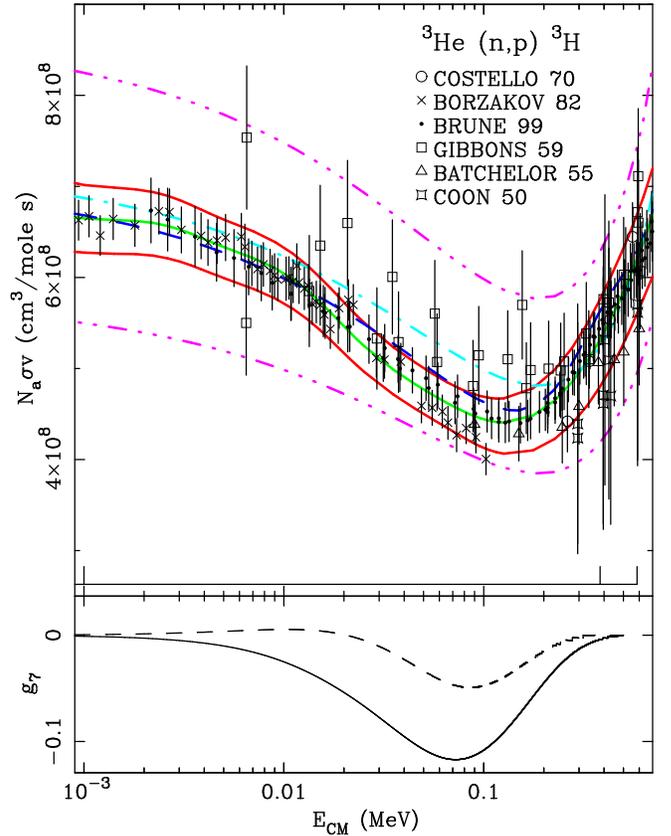}}
\caption{$S$ factor data and fits for $^3{\rm He}(n,p) t$.}
\label{fig:he3np}
\end{figure}

The data set for this reaction consists of cross sections for both the
forward
(Refs. \cite{he3np-borzakov,he3np-costello,he3np-coon,he3np-batchelor})
and reverse (Ref. \cite{he3np-brune,he3np-gibbonsmacklin}) processes.
We used reverse data that had been converted to forward cross sections
through exact detailed balance relations, along with direct data.  The
SKM fit included numbers from Alfimenkov \cite{he3np-alfimenkov},
which we did not obtain, but the important difference between our
analysis and that of SKM is that they excluded the very precise
measurement of Borzakov {\it et al.}  \cite{he3np-borzakov}, which
owes its small quoted uncertainty to a normalization from $^6{\rm
Li}(n,\alpha)^3{\rm He}$ data intended for metrological use
\cite{li6na-gayther,li6na-lamaze}.  SKM excluded this data set because
it decreases with energy more quickly than the other data, ending
lower than any of them at 100 keV.  This, combined with the small
quoted errors, would have forced their second-order polynomial fit
below the data at higher energies.  However, this is not a problem for
our piecewise spline approach, which decouples errors arising from
different experiments at different energies.  The extensive and
much-needed data which Brune {\it et al.} \cite{he3np-brune} published
recently for this cross section also agree more closely with the
Borzakov {\it et al.} than with SKM.

\subsubsection{$t(d,n)^4{\rm He}$}

\begin{figure}
\centerline{\epsfig{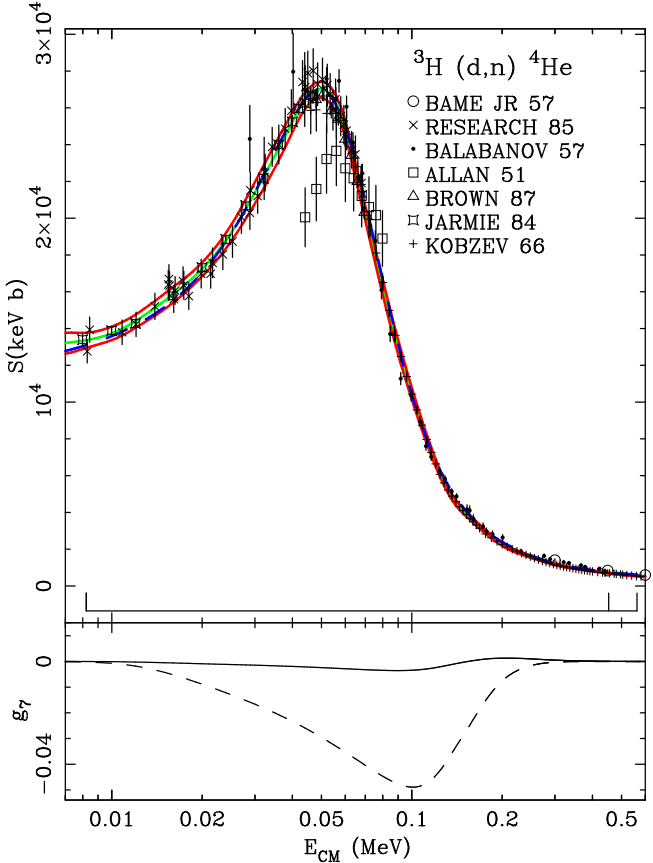}}
\caption{$S$ factor data and fits for $t(d,n)^4{\rm He}$.}
\label{fig:tdn}
\end{figure}

The definitive measurement of the cross section for this process, up
to 70 keV, is by the Los Alamos group \cite{tdn-bjh,tdn-jbh}, and it
has a quoted normalization error of 1.4\%.  Other measurements in our
database come from
Refs. \cite{dd-first,tdn-bame,tdn-allan,tdn-balabanov,tdn-kobzev}.
As SKM point out, there are no modern experiments on the high-energy
side of the resonance.  However, this reaction does not contribute
noticeably to the uncertainty in D or $^7$Li yields.

\subsubsection{$^3{\rm He}(d,p)^4{\rm He}$}
\label{sec:he3dp}

\begin{figure}
\centerline{\epsfig{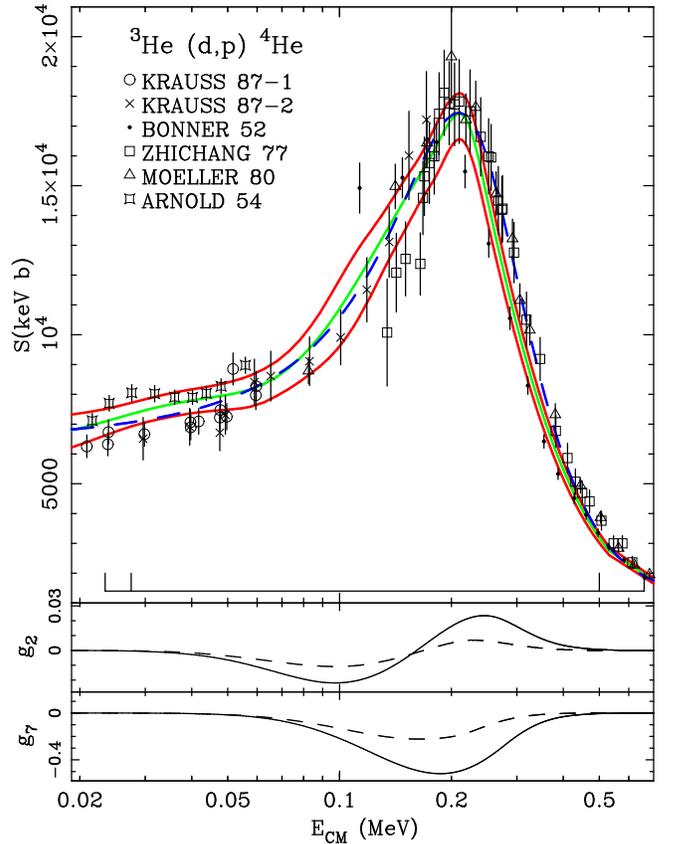}}
\caption{$S$ factor data and fits for $^3{\rm He}(d,p)^4{\rm He}$.}
\label{fig:he3dp}
\end{figure}

There is considerable scatter in the data for this process, as
reflected in the very low probability of the $\chi^2$ statistic.  In
the case of the Bonner {\it et al.}  measurement \cite{he3dp-bonner},
this may be attributable to energy straggling in the gas target's Al
window.  In any case, there is little else to say here except to point
out that our piecewise spline seems to represent the data about as
well as the SKM $R$-matrix fit.  This reaction contributes a large
portion of the uncertainty in the $^3$He yield.  Other data used for
this reaction are from Refs.  \cite{dd-krauss} and
\cite{he3dp-zhichang,he3dp-moeller,he3dp-arnold}.  The low-energy data
of Schr\"oder {\it et al.} \cite{he3dp-schroeder}, intended to probe
electron screening, were omitted here because they did not measure the
absolute cross section.

\subsubsection{$^3{\rm He}(\alpha,\gamma)^7{\rm Be}$}

\label{sec:he3ag}

\begin{figure}
\centerline{\epsfig{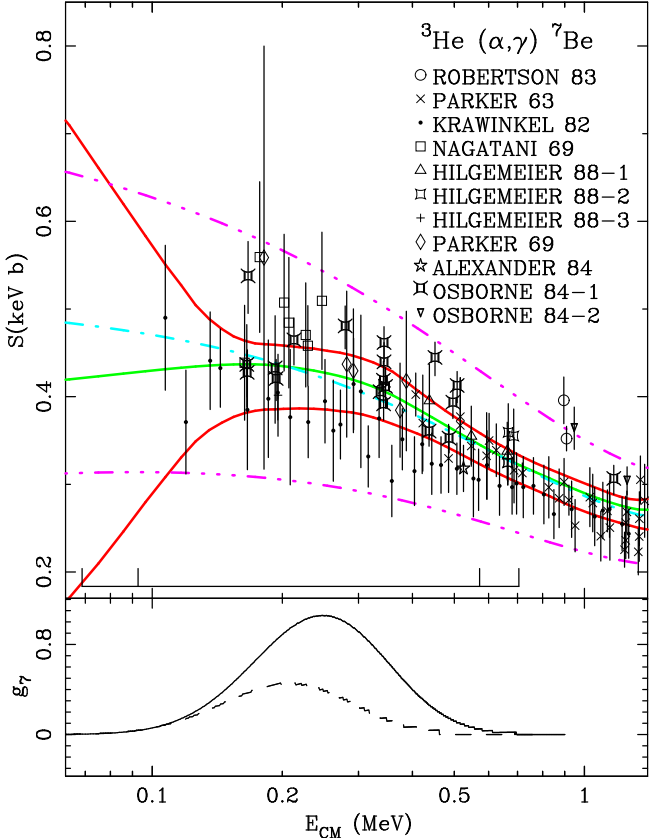}}
\caption{$S$-factor data and fits for $^3{\rm He}(\alpha,\gamma)^7{\rm Be}$.
Note that the data of Kr\"awinkel {\it et al.} \protect\cite{he3ag-kraewinkel}
have been renormalized as suggested by Hilgemeier {\it et al.}
\protect\cite{he3ag-hilgemeier}.}
\label{fig:he3ag}
\end{figure}

The apparent discrepancy at $2\sigma$ between experiments that observe
capture gamma rays from this process (Refs.
\cite{he3ag-kraewinkel,he3ag-hilgemeier,he3ag-parker,he3ag-nagatani,he3ag-alexander,he3ag-osborne})
and those that observe photons from the $^7$Be decay
(Refs.\cite{he3ag-osborne,he3ag-robertson,he3ag-volk}) is well-known
from discussions of the solar neutrino problem \cite{adelberger}.  Our
direct approach to the data makes no use of the (well-understood)
theory of this process, and the activation technique has only been
used at energies too high to affect big-bang nucleosynthesis (although
the Volk {\it et al.} \cite{he3ag-volk} results are quoted as
extrapolated $S(0)$ values only).  However, the energy dependence of
this reaction is sufficiently well-determined from the capture photon
experiments to justify the use of an ``activation method'' curve
obtained by renormalizing the fit of the capture-photon-derived cross
sections to match the activation points.

After the main run, we re-ran our Monte Carlo sampling, renormalizing
the $^3{\rm He}(\alpha,\gamma)^7{\rm Be}$ cross section from the
capture-photon measurements by the mean of the activation
measurements, and drawing the renormalization from a Gaussian
distribution based on the variance of the activation measurements.
For this purpose, we use the reference of all cross sections to zero
energy found in Adelberger {\it et al.}  \cite{adelberger}.  The
effect of this discrepancy (amounting to a systematic shift of 13\%)
is easy to understand; all $^7$Li produced at high $\Omega_B$ comes
from this reaction, so changes in its rate result directly in changes
of the final $^7$Li yield.  The result is a shift of 11\% in our
confidence limits for $^7$Li, a significant fraction of the widths of
these limits, at high $\Omega_B$.  (See below.)  If the activation
measurements are correct, this would exacerbate the problem of $^7$Li
depletion in halo stars --- a point which has not previously arisen in
discussions of this problem because it represents a much smaller
fraction of the widths of the SKM $^7$Li limits, and because SKM
dropped the activation data from their evaluation altogether.

\subsubsection{$ t(\alpha,\gamma)^7{\rm Li}$}

\begin{figure}
\centerline{\epsfig{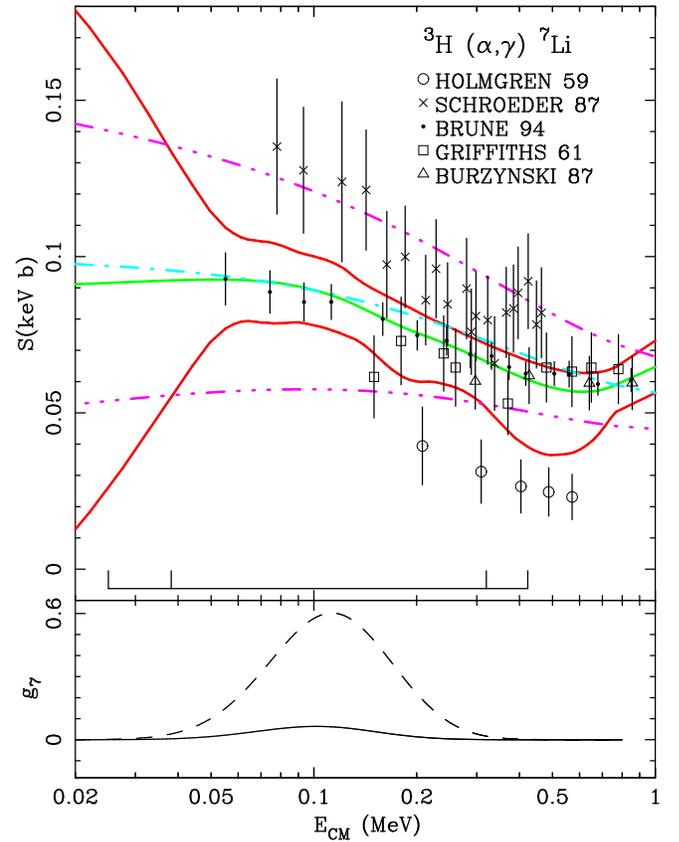}}
\caption{$S$ factor data and fits for $ t(\alpha,\gamma)^7{\rm Li}$.}
\label{fig:tag}
\end{figure}

A precise new measurement of the cross section for this process has
been completed by Brune {\it et al.} \cite{tag-brune} in the time
since the publication of SKM.  Some subsequent calculations have
incorporated the reaction rate derived from that measurement by its
authors ({\it e.g.,} Copi {\it et al.} \cite{cst}), using as the
uncertainty the 6\% uncertainty in the experiment's cross section
normalization.  Krauss and Romanelli \cite{kr} pointed out that it is
not always the best policy to base calculations on only the most
recent measurement, and we agree.  The previous data, Refs.
\cite{tag-holmgren,tag-schroeder,tag-griffiths,tag-burzynski}, do show
a great deal of scatter, and our best-fit curve tends to follow the
more-precise Brune {\it et al.} data, as it should.  Relative to the
SKM yields, this has little effect at high baryon density, but it
decreases the $^7$Li yield slightly at very low baryon density, where
this process makes most of the $^7$Li.  We have omitted the
Coulomb-breakup measurement of Utsunomiya {\it et al.}
\cite{tag-utsunomiya} because the Coulomb-breakup process is not
completely understood for this reaction (as discussed in SKM) and
because the cross section energy dependence derived from this method
disagrees with the Brune {\it et al.}  data.  We note that the problem
with normalization of the Schr\"oder {\it et al.}
\cite{tag-schroeder} and Griffiths {\it et al.} \cite{tag-griffiths}
data sets mentioned by SKM seems to have gone away in the intervening
time \cite{tpg-hahn}.

\subsubsection{$^7{\rm Be}(n,p)^7{\rm Li}$}

\begin{figure}
\centerline{\epsfig{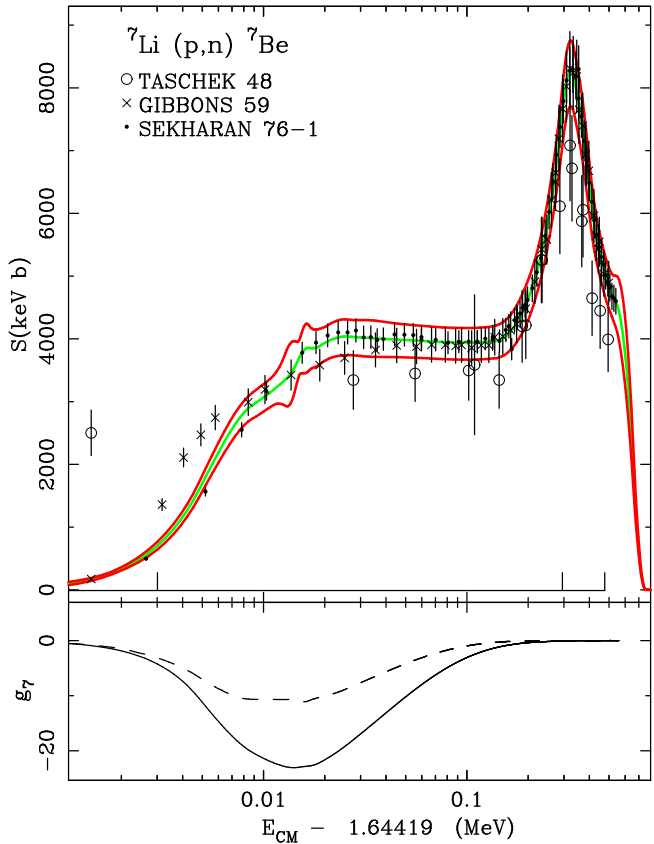}}
\caption{$S$ factor data and fits for $^7{\rm Li}(p,n)^7{\rm Be}$.}
\label{fig:li7pn}
\end{figure}

We fitted the cross section for the reverse of this process, since
there are no direct data in the energy region of interest for BBN.
This may not be the ideal choice, since the curve has a very large
second derivative just above threshold.  We fit data for $^7{\rm
Li}(n,p)^7{\rm Be}$ from Sekharan {\it et al.}  \cite{li7pn-sekharan},
Taschek and Hemmendinger \cite{li7pn-taschek}, and Gibbons and Macklin
\cite{he3np-gibbonsmacklin}, which extend from threshold to well past
the lowest-energy resonance.  We also cut off these data sets at about
700 keV above threshold (where they no longer affect BBN yields), so
that only data in the critical range affect the $\chi^2$ calculations.
Although this reaction contributes very little to our error budget
(see Sec. \ref{sec:results} below), it is important to recognize the
scarcity of data and the lack of detailed error analysis in the
original sources.

\subsubsection{$^7{\rm Li}(p,\alpha)^4{\rm He}$}

\begin{figure}
\centerline{\epsfig{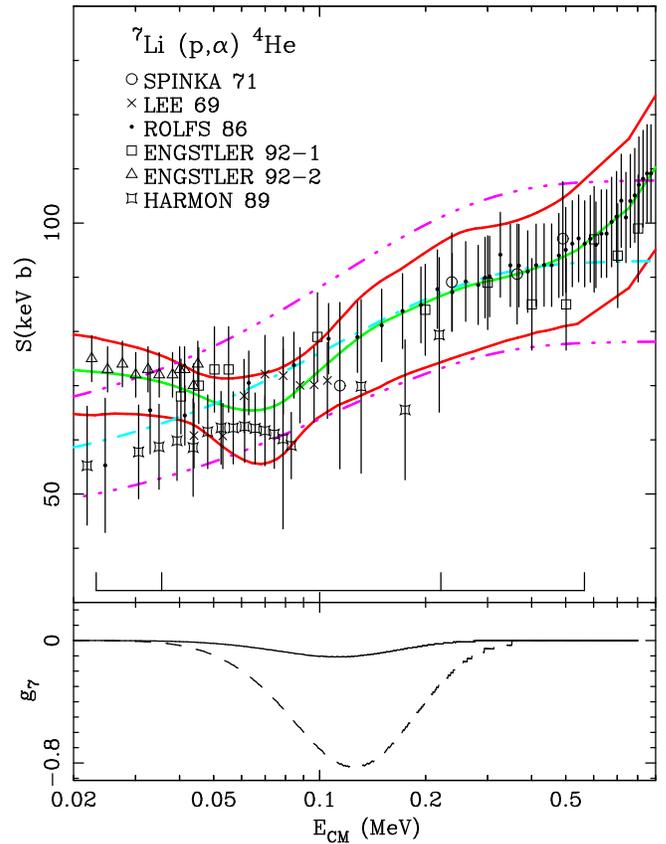}}
\caption{$S$-factor data and fits for $^7{\rm Li}(p,\alpha)^4{\rm He}$.}
\label{fig:li7pa}
\end{figure}

The low-energy cross section for this process is represented by a
small number of measurements, Refs.
\cite{li7pa-spinka,li7pa-chul,li7pa-rolfs,li7pa-engstler,li7pa-harmon},
and is determined over most of the energy interval of interest to us
by the data of Rolfs {\it et al.} \cite{li7pa-rolfs}.  We have kept
the data of Engstler {\it et al.} \cite{li7pa-engstler} below 50 keV
in our fit, even though the slight rise in this data set relative to
that of Harmon \cite{li7pa-harmon} has been attributed to electron
screening effects.  (The data of Harmon were normalized to the $^6{\rm
Li}(p,\alpha)^3{\rm He}$ cross section in such a way that they have
been ``corrected'' for screening to some extent
\cite{li7pa-engstler}.)  The fact that our $S$-factor curves follow
the small error bars of the low-energy Engstler {\it et al.} data is
not too distressing, since the sensitivity function for lithium
production for this reaction is very small below 40 keV, and a true
correction for screening would require a thorough theoretical
treatment, {\it e.g.,} $R$-matrix techniques with a direct reaction
mechanism.  It is not clear that simply extrapolating fits to simple
functions from higher energies is valid, or that the data of Harmon
{\it et al.} reflect a true correction for screening.

\subsection{Recommended Rates and Errors}

A disadvantage of our method, relative to earlier efforts, is that it
requires the whole nuclear database and a larger amount of computer
time, as well as a significant modification of existing BBN code.
Reaction rate uncertainties are also harder to quote than in the SKM
prescription, since uncertainties at different temperatures are
neither completely correlated nor completely uncorrelated.  In any
case, the rates produced by our code are not suitable for any use
significantly different from the standard BBN calculation, and our
method is specific to the BBN context, where only a few cross sections
--- well-represented at the right energies by available data --- are
needed.  We note, in particular, that the ``flaring'' of our piecewise
polynomial fits at high and low energy does not contain any physical
information at all, but only the fact that polynomial interpolations
always blow up beyond the limits of the data they were fitted to.  For
general use (especially outside the BBN energy region), we suggest
using rates from a more general compilation, such as the NACRE
compilation \cite{nacre}, intended to succeed the Caughlin and Fowler
\cite{cf} charged-particle reaction rates.

\section{results}
\label{sec:results}

\begin{figure}
\centerline{\epsfxsize=10cm \epsfig{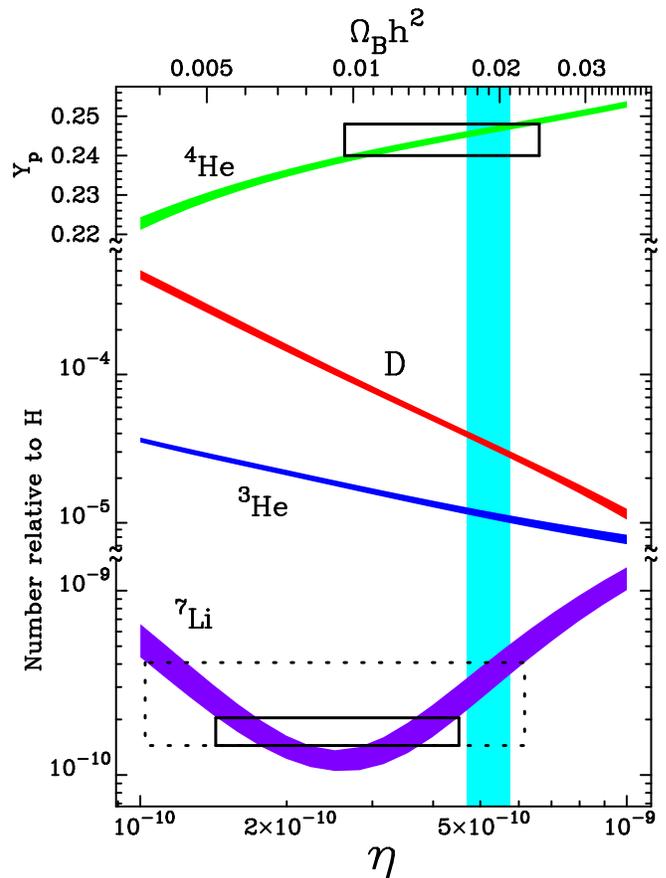}}
\caption{Summary of the 95\% confidence intervals for the BBN
predictions for D, $^3$He, $^4$He and $^7$Li.  The $^4$He uncertainty
comes from Ref.~\protect\cite{lopezturner98}.  Boxes indicate 95\% cl
abundances from observation \protect\cite{bono,it98,oss97}.  The
vertical band indicates our 95\% cl baryon density inferred from the
Burles and Tytler ~\protect\cite{burlestytler98a,burlestytler98b}
deuterium observations.}
\label{fig:etaplot}
\end{figure}

\begin{figure}
\centerline{\epsfig{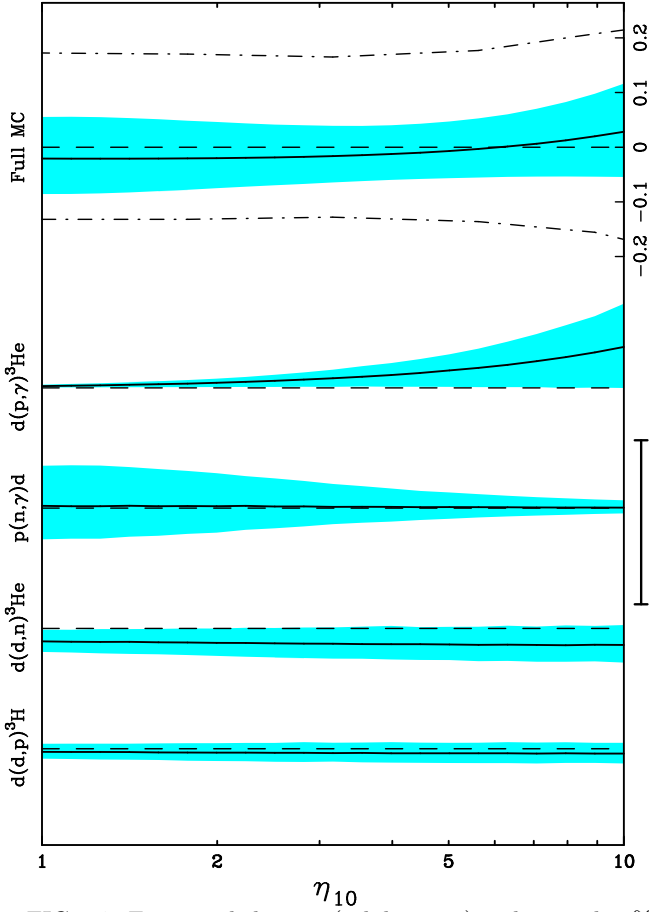}}
\caption{Fractional changes (solid curves) and revised 95\% confidence
limits (shaded regions) in D/H from our full Monte Carlo calculation
and from Monte Carlo studies of individual reactions, relative to the
SKM yields and errors (dashed lines).  For comparison, the vertical
bar at the side of the figure indicates the $1\sigma$ uncertainty in
the Burles and Tytler D/H measurements \protect\cite{burlestytler98a}.
All curves are on the same scale.}
\label{fig:d-errors}
\end{figure}

In discussing the results of applying our prescription, we concentrate
on the predictions of $^7$Li and D yields.  On the one hand, the
observational status of $^3$He is not such as to motivate precise
comparisons with the calculation.  On the other, the errors in $^4$He,
especially at higher values of $\Omega_B$, are dominated by the
uncertainty in the weak coupling constant and by uncertainties in
calculating the matrix elements for the weak processes.  These errors
and theoretical uncertainties have been analyzed exhaustively by Lopez
and Turner \cite{lopezturner98}, and we have not included all the
apparatus of their treatment in our BBN code; we take their results
for this nuclide to be definitive.

\begin{figure}
\centerline{\epsfig{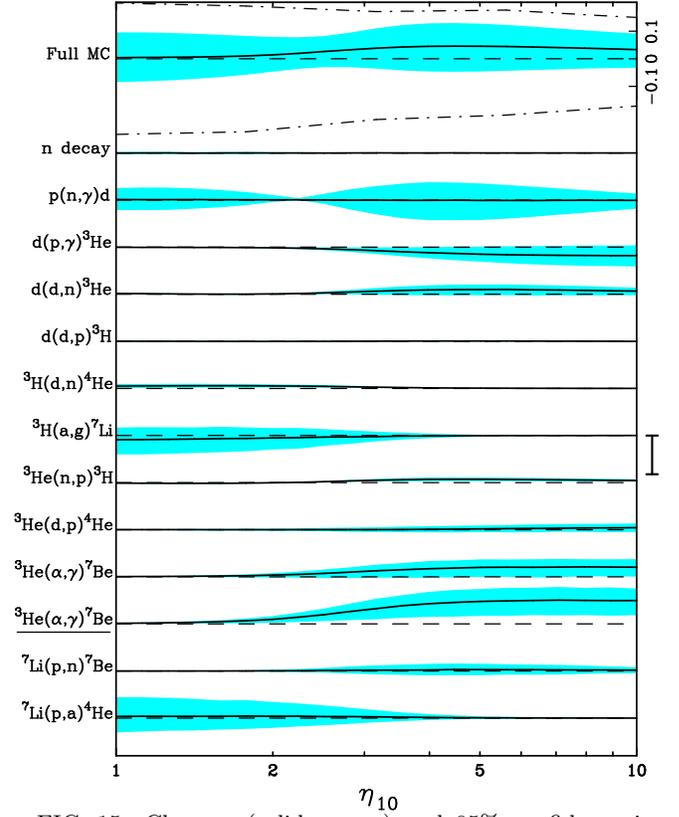}}
\caption{Changes (solid curves) and 95\% confidence intervals (shaded
regions) in $\log(^7{\rm Li/H})$ from our full Monte Carlo calculation
and from Monte Carlo studies of individual reactions, relative to the
SKM yields and errors (dashed lines).  The vertical bar on the side
indicates $1\sigma$ error in the Spite-plateau abundance
\protect\cite{bono}.  The underlined reaction label indicates the
individual-variation result using the alternate rate for $^3{\rm
He}(\alpha,\gamma)^7{\rm Be}$, as described in the text.  All curves
are shown on the same scale.}
\label{fig:li7-errors}
\end{figure}

Our most important results, apparent from Figs. \ref{fig:etaplot},
\ref{fig:d-errors}, and \ref{fig:li7-errors}, are reductions by
factors of up to three in the width of the 95\% confidence intervals
for both the $^7$Li and D yields relative to SKM.  The median values
of our yields are almost identical to those obtained from the SKM
rates, the difference being well within our estimated errors.  These
small changes can generally not be attributed to any one reaction, but
to some nonlinear addition of changes from several reactions, as
indicated in Fig. \ref{fig:li7-errors}.

We did not expect a huge change either in the size of the error
estimates or in the calculated most-likely BBN yields.  We did expect
some reduction in the error estimate because we did not go out of our
way to be conservative.  The reduction in the size of our error
estimates is largely a side-effect of our method of handling the
nuclear data, chosen as a way of relating uncertainties in the
calculated yields to uncertainties in the nuclear data.

We also studied the error contributions of individual reactions by
setting all rates but one to their SKM values, and applying our Monte
Carlo technique to that reaction alone.  As shown in
Figs. \ref{fig:d-errors} and \ref{fig:li7-errors}, this indicates the
relative importance of each reaction as a function of baryon density
(much as in the figures of Krauss and Romanelli \cite{kr}).  This
shows exactly which reaction rates need improvement to reduce the
errors on BBN yields.  In turn, the knowledge of which cross sections
need improvement can be combined with the sensitivity functions of
Sec. \ref{sec:data} and Figs. \ref{fig:png} through \ref{fig:li7pa} to
indicate the specific energies at which they need improvement.  The
``most wanted'' for the deuterium abundance are, from most to least
important at $\Omega_Bh^2=0.019$: $d(d,n)^3{\rm He}$ above 100 keV;
$d(p,\gamma)^3{\rm He}$ everywhere; $d(d,p)^3{\rm H}$ above 100 keV,
and $p(n,\gamma)d$ at 30--200 keV.  For $^7$Li, the leading
contributions to the uncertainty at $\Omega_Bh^2=0.019$ are from
$p(n,\gamma)d$ at 20--150 keV, $^3{\rm He}(\alpha,\gamma)^7{\rm Be}$
at 150-375 keV (and overall normalization), $d(p,\gamma)^3{\rm He}$
everywhere, and $d(d,n)^3{\rm He}$ above 100 keV.  At
$\Omega_Bh^2=0.009$, the uncertainty in $^7$Li comes mainly from
$^3{\rm H}(\alpha,\gamma)^7{\rm Li}$ and $^7{\rm Li}(p,\alpha)^4{\rm
He}$.

Taking this list one step further, we have generated fake data for
some of these reactions (following the best-fit curve) and placed it
in our database.  In each case, fake data were placed on twenty
evenly-spaced intervals between 5 and 500 keV center-of-mass.  We then
re-did the single-reaction Monte Carlo for that reaction, and reduced
the size of the normalization error on the fake data set until the
uncertainty estimate due to that reaction was reduced by half.  The
sizes of the normalization uncertainties required by this criterion
are given in Table \ref{tab:fake}.  We assumed that unshared errors
can be made arbitrarily small, and left them out.  Similar fake data
sets modeling proposed experiments could be used to determine what
effect they would have on the BBN error estimates in our formalism.

\begin{table}
\caption{Size of normalization error necessary in a new data set (of
twenty points evenly-spaced from 5 to 500 keV) for its addition to our
database to reduce the single-reaction error estimate on the indicated
BBN yields by half at $\Omega_Bh^2=0.019$.}
\begin{tabular}{cdc}
Reaction & Desired error in cross section & BBN product \\
\tableline
$d(p,\gamma)^3{\rm He}$  & 3.2\% & D \\
`$d(d,n)^3{\rm He}$ & 2.0\% & D \\
$d(d,p)^3{\rm H}$ & 1.6\% & D \\
$^3{\rm He}(\alpha,\gamma)^7{\rm Be}$ & 2.3\% & $^7$Li \\
\end{tabular}
\label{tab:fake}
\end{table}

\begin{figure}
\centerline{\epsfig{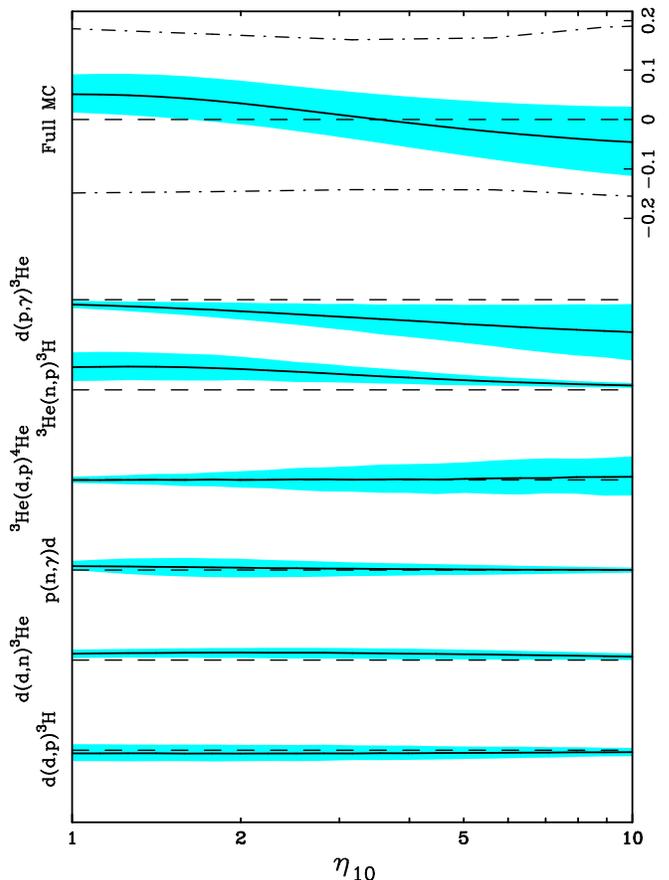}}
\caption{Fractional changes (solid curves) and 95\% confidence
intervals (shaded regions) in $^3$He/H from our full Monte Carlo
calculation and from Monte Carlo studies of individual reactions,
relative to the SKM yields and errors (dashed lines).  All curves are
shown on the same scale.  }
\label{fig:he3-errors}
\end{figure}

\begin{figure}
\centerline{\epsfig{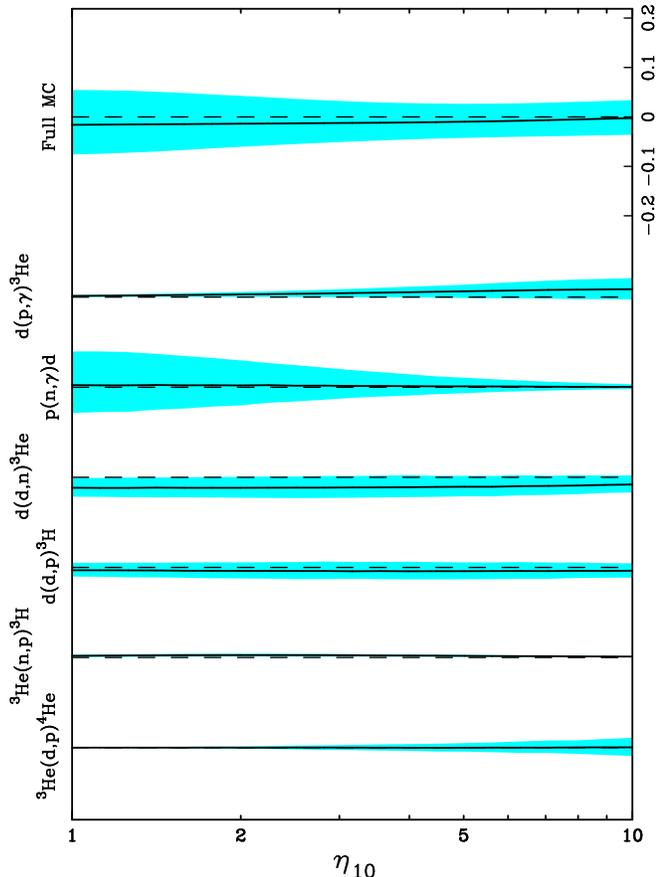}}
\caption{Fractional changes (solid curves) and 95\% confidence
intervals (shaded regions) in ($^3$He+D)/H from our full Monte Carlo
calculation and from Monte Carlo studies of individual reactions,
relative to the SKM yields (dashed lines).  All curves are
shown on the same scale.  }
\label{fig:dhe3-errors}
\end{figure}

While the $^3$He chemical evolution and abundance measurements are too
uncertain to motivate high-precision comparisons with the calculation
\cite{st98,schramm97,rood98,balser98}, we have also examined the
results of our calculations for this nuclide and for its reaction-rate
dependences.  The results are indicated in Fig. \ref{fig:he3-errors}.
The slope of the dependence of this abundance on baryon density has
changed relative to the SKM rates.  At low $\Omega_B$, this reflects
the reduced rate of $^3{\rm He}(n,p)^3{\rm H}$ at about 100 keV in our
calculations (a result of our fit emphasizing more precise
measurements, but reinforced by very recent measurements).  At high
$\Omega_B$, this reflects the reduced rate of $d(p,\gamma)^3{\rm He}$
indicated by recent improved measurements.  These effects cancel in
the middle of the Copi {\it et al.}  \cite{cst} concordance interval,
so that the disagreement with SKM is not serious in the likely range
for $\Omega_B$.  Since much of the post-big-bang evolution of D/H and
$^3$He/H is expected to consist of the burning of deuterium into
$^3$He, the sum of these two number densities is often considered in
comparing them to the BBN predictions.  Therefore, we also show the
limits on this sum from our Monte Carlo in Fig. \ref{fig:dhe3-errors}.

\section{conclusions}
\label{sec:conclusions}

In summary, we have applied a new Monte Carlo approach to the use of
nuclear data in big-bang nucleosynthesis calculations.  This approach
has the virtues of coupling the results of BBN calculations and their
error estimates closely to the available nuclear data, of explicitly
handling correlated errors in the data set, of allowing easy use of
new data, and of taking some of the data-evaluation process out of
human hands.  We have also abandoned the explicit conservatism of the
previous ``industry-standard'' error estimation, but the chief virtue
of our method is the close coupling of the BBN calculation to the
nuclear data, particularly with regard to incorporating new data.

Application of our method has resulted in a reduction in the estimated
uncertainty in the BBN calculation.  The old estimates of the
uncertainties were larger than the quoted uncertainties in recent
astronomical observations of BBN nuclides, so that uncertainties in
the nuclear inputs dominated inferences from individual observations.
Given our prescription, the uncertainties in the calculation are once
again smaller than the quoted errors on any of the current
observations, and this strengthens the constraints which can be
placed on BBN.  The constraints that can be derived by applying our
calculation to current observations have been discussed in a previous
paper \cite{us-prl}.

The most important result of our earlier paper is that useful
inferences can now be made where nuclear uncertainties formerly
precluded any strong conclusions.  An example is the question of
whether observed lithium abundances are consistent with low deuterium
observations, in the absence of lithium depletion on the Spite
plateau.  The conservative uncertainty estimates would also not allow
any determination of the baryon density based on BBN to better than
about 10\%; our prescription reduces the uncertainty in this quantity
so that even given observational errors of a few percent in deuterium,
the uncertainty on the baryon density would be dominated by
astronomical observations.

We have subsequently found that the calculated BBN yields in Ref.
\cite{us-prl} suffer from a programming error which we have now
corrected.  The error resulted in {\it overestimates} of the
uncertainties by up to a factor of about 1.5 in Figure 1 of that
paper.  Discussion in the earlier paper concentrated on uncertainties
at $\Omega_Bh^2=0.019$; at that baryon density, correction of the
programming error reduces the estimated uncertainty on D/H by a factor
of 1.6 and the estimated uncertainty on $^7$Li/H by a factor of 1.2.
Consequences of this error for cosmological implications are
relatively unimportant, because the uncertainty estimates on our
preliminary calculation are already smaller than those on astronomical
observations.

With regard to the nuclear data, we point out that any lingering
problems with the nuclear database will probably have to be settled by
new experiments.  The low-energy $p(n,\gamma)d$ cross section is an
important gap; the $^3{\rm He}(d,p)^4{\rm He}$ cross section is a less
important gap, but it is clearly the case with the worst systematic
problems in our nuclear database.  Further reactions that contribute
to lithium and deuterium yield uncertainties are listed in
Sec. \ref{sec:results}.  The importance of improving various
components of the nuclear database relative to improving the
astronomical observations can be seen by examining Figs.
\ref{fig:d-errors}--\ref{fig:dhe3-errors}.

It is possible to improve on our work.  We have explicitly tried to
work very directly with the nuclear inputs, and we have therefore
avoided theoretical modelling, which is not essential in the face of
such copious data.  With more theoretical inputs --- for example,
radiative-capture models and multi-channel $R$-matrix fits --- one
could include much more data, from higher energies and from scattering
channels, and introduce some cross-talk between reaction channels.
Solely in terms of gathering and processing data, this would be a much
larger undertaking than what we have done.  It could probably not be
done piecemeal, because important theoretical parameters often have to
be determined by examining data in several reaction and scattering
channels, and over a wide range in energy.  Depending on
implementation, theoretical inputs may also present a more difficult
problem in terms of propagating errors through to BBN reaction rates.

Improvement may also be possible in our handling of correlated
normalization errors.  We have also avoided any fancier error
estimation than applying quoted errors in a well-defined way.  In
particulary, we have not fit with floating normalizations because we
were wary of altering the most important pieces of information for
BBN.  In place of floating-normalization methods, we have introduced a
Monte Carlo method for making families of smooth curves to
characterize data with normalizations that vary in the expected way.
Although we do believe that this characterizes the uncertainties in
the database in a reasonable way, it is worth noting that the
individual curve fitted through each realization of the data does not
recognize the presence of correlated errors.

In conclusion, our results indicate that our method of coupling BBN
error estimates to the nuclear data may be fruitful not only for
providing a useful and unambiguous prescription for such error
estimates, but also for making comparison between light-element
abundances and BBN calculations more meaningful.  While our
prescription for handling the nuclear data is not unique, it is
simple, repeatable, and direct.  Such a method is needed for the
program of ``precision cosmology.''  We hope that our proposal will
result at least in a more critical stance toward uncertainty estimates
in BBN, and perhaps improved prescriptions that incorporate the better
qualities of our method.

\acknowledgments

We thank Carl Brune for providing numerical cross section data, and we
thank Michael Turner and Jim Truran for helpful discussions.  This
work was initiated with David N. Schramm.

%\bibliographystyle{prsty}
%\bibliography{prd}

\end{document}